\documentclass[journal,twoside,web]{IEEEtran}
\usepackage[utf8]{inputenc}
\usepackage{cite}
\usepackage{graphicx}
\usepackage{textcomp}

\usepackage{stackengine}
\newcommand\barbelow[1]{\stackunder[1.2pt]{$#1$}{\rule{.8ex}{.075ex}}}
\usepackage{float}

\usepackage{amsmath}
\usepackage{nccmath}
\usepackage{amsfonts}
\usepackage{amssymb}
\usepackage{relsize}

\usepackage{multirow}
\usepackage{siunitx}
\usepackage{stackengine}
\usepackage{cite}
\usepackage{float}
\usepackage{ifthen}
\usepackage{etoolbox}
\usepackage{soul}
\usepackage{array}
\usepackage{psfrag}
\usepackage{epsfig}

\usepackage{pifont}
\usepackage{wasysym}

\usepackage[]{algorithmicx}
\usepackage{algpseudocode}
\usepackage{algorithm}
\usepackage{booktabs}

\usepackage[caption=false, font=footnotesize]{subfig}
\algnewcommand{\LeftComment}[1]{\Statex \(\triangleright\) #1}

\usepackage[symbol]{footmisc}

\usepackage{cleveref}

\usepackage{tikz}
\usetikzlibrary{calc,shapes,decorations,decorations.text, mindmap,shadings,patterns,matrix,arrows,intersections,automata,backgrounds,shadows,arrows.meta}

\newcommand{\twopartdef}[4]
{
	\left\{
	\begin{array}{ll}
		#1 & \mbox{if } #2 \\
		#3 & \mbox{if } #4
	\end{array}
	\right.
}

\usepackage{framed} 
\usepackage{multicol} 
\usepackage{nomencl} 
\makenomenclature
\renewcommand*\nompreamble{\begin{multicols}{2}}
\renewcommand*\nompostamble{\end{multicols}}

\usepackage{etoolbox}
\renewcommand\nomgroup[1]{%
  \item[\itshape
  \ifstrequal{#1}{A}{Sets}{%
  \ifstrequal{#1}{B}{Variables}{%
  \ifstrequal{#1}{C}{Parameters}{%
  \ifstrequal{#1}{D}{Abbreviations}{
  \ifstrequal{#1}{E}{Tariff nomenclature}{}}}}}%
]}

\def\BibTeX{{\rm B\kern-.05em{\sc i\kern-.025em b}\kern-.08em
    T\kern-.1667em\lower.7ex\hbox{E}\kern-.125emX}}
\begin{document}
\title{Probabilistic assessment of the impact of flexible loads under network tariffs in low voltage distribution networks}
\author{Donald~Azuatalam,~\IEEEmembership{Graduate Student Member,~IEEE,}
	Archie~C.~Chapman,~\IEEEmembership{Member,~IEEE,}
	and~Gregor~Verbi\v{c},~\IEEEmembership{Senior Member,~IEEE}
\thanks{Donald Azuatalam, Archie C. Chapman and Gregor Verbi\v{c}, are with the School of Electrical and Information Engineering, The University of Sydney, Sydney, New South Wales, Australia. E-mail: donald.azuatalam@sydney.edu.au, archie.chapman@sydney.edu.au, gregor.verbic@sydney.edu.au.}}

\maketitle

\begin{abstract}
Given the historically static nature of low-voltage networks, distribution network companies do not possess tools for dealing with an increasingly variable demand due to the high penetration of distributed energy resources (DER). Within this context, this paper proposes a probabilistic framework for tariff design that minimises the impact of DER on network performance, stabilise network company revenue, and improves the equity of network costs allocation. To address the issue of the lack of customers’ response, we also show how DER-specific tariffs can be complemented with an automated home energy management system (HEMS) that reduces peak demand while retaining the desired comfort level. The proposed framework comprises a nonparametric Bayesian model which statistically generates synthetic load and PV traces, a hot-water-use statistical model, a novel HEMS to schedule customers’ controllable devices, and a probabilistic power-flow model. Test cases using both energy- and demand-based network tariffs show that flat tariffs with a peak demand component reduce the customers’ cost, and alleviate network constraints. This demonstrates, first, the efficacy of the proposed tool for the development of tariffs that are beneficial for networks with a high DER penetration, and second, how customers’ HEM systems can be part of the solution.
\end{abstract}



\begin{table*}[!t] \label{nomen}
   \begin{framed}
     \printnomenclature
   \end{framed}
\end{table*}

\nomenclature[A]{$\mathcal{D}$}{Set of days, $d \in \mathcal{D}$ in a year, $\mathcal{D} = \{1,...,365\}$}
\nomenclature[A]{$\mathcal{D}'$}{Set of days, $d' \in \mathcal{D}'$ in a month, $\mathcal{D}' \subset \mathcal{D}$}
\nomenclature[A]{$\mathcal{H}$}{Set of half-hour time-slots, $h \in \mathcal{H}$ in a day, \\ $\mathcal{H} = \{1,...,48\}$}
\nomenclature[A]{$\mathcal{M}$}{Set of months, $m \in \mathcal{M}$ in a year, $\mathcal{M} = \{1,...,12\}$}

\nomenclature[B]{$\hat{p}$}{Dummy variable for modelling demand-based tariffs}
\nomenclature[B]{$p^\mathrm{g+/-}$}{Power flowing from/to grid}
\nomenclature[B]{$p^\mathrm{b+/-}$}{Battery charge/discharge power}
\nomenclature[B]{$e^\mathrm{b}$}{Battery state of charge}
\nomenclature[B]{$s^\mathrm{b}$}{Battery charging status (0: discharge, 1: charge)}
\nomenclature[B]{$d^\mathrm{g}$}{direction of grid power flow (0: demand to grid, 1: grid to demand)}

\nomenclature[C]{$\eta^\mathrm{b+/-}$}{Battery charging/discharging efficiency}
\nomenclature[C]{$\bar{p}^\mathrm{b+/-}$}{Maximum battery charge/discharge power}
\nomenclature[C]{$\eta^\mathrm{b+/-}$}{Battery charging/discharging efficiency}
\nomenclature[C]{$\bar{e}^\mathrm{b}$}{Battery maximum state of charge}
\nomenclature[C]{$\barbelow{e}^\mathrm{b}$}{Battery minimum state of charge}
\nomenclature[C]{$p^\mathrm{pv}$}{Power from solar PV}
\nomenclature[C]{$\Delta h$}{Half hourly time steps}
\nomenclature[C]{$p^\mathrm{d}$}{Total customer demand}
\nomenclature[C]{$p^\mathrm{res}$}{Net demand}
\nomenclature[C]{$\bar{p}^\mathrm{g}$}{Maximum power taken from/to grid}

\nomenclature[D]{LV}{Low voltage}
\nomenclature[D]{PV}{Photovoltaic}
\nomenclature[D]{BESS}{Battery energy storage system}
\nomenclature[D]{DER}{Distributed energy resources}
\nomenclature[D]{DNSP}{Distribution network service provider}
\nomenclature[D]{FiT}{Feed in tariff}
\nomenclature[D]{ToU}{Time of use}
\nomenclature[D]{MILP}{Mixed integer linear programming}
\nomenclature[D]{HEMS}{Home energy management system}
\nomenclature[D]{EWH}{Electric water heater}

\nomenclature[E]{$T^\mathrm{flt}$}{Flat energy charge} 
\nomenclature[E]{$T^\mathrm{tou}$}{Time-of-use energy charge} 
\nomenclature[E]{$T^\mathrm{fix}$}{Fixed daily charge}
\nomenclature[E]{$T^\mathrm{fit}$}{Feed-in-tariff (FiT)}
\nomenclature[E]{$p^\mathrm{pk}$}{monthly peak}
\nomenclature[E]{$T^\mathrm{pk}$}{Monthly Peak demand charge}


\section{Introduction}


Investment in customer-owned PV-battery systems is growing rapidly across the globe, as they become cost-effective in certain jurisdictions. 
For example, the total installed capacity of residential PV-battery systems in Australia is projected to increase from 5~\si{GW} in 2017 to 19.7~\si{GW} in 2037~\cite{aemosmall,aemosolar}. In Germany, the total installed capacity of PV systems alone currently stands at 43~\si{GW}, and projected to increase to 150~\si{GW} by 2050~\cite{isefraunhofer,wirth2018recent}; 
while battery storage systems are expected to follow suit, with currently 100,000 installations (approx. 6~\si{GWh}) and projections for this to double within the next two years~\cite{bswsolar}.

The trend towards more residential PV-battery systems is being driven by two major factors.
On one hand, average household electricity prices in OECD countries have increased by over 33\% between 2006 and 2017 (using purchasing power parity). 
In particular, in Australia and Germany, prices have risen to about 20.4 and 39.17~\si{US. c/kWh}, respectively, from roughly 12.52~\si{US. c/kWh} (in Australia) and 20.83~\si{US. c/kWh} (in Germany) in the year 2006~\cite{energytaxes}; while feed-in-tariff (FiT) rates for PV generation have been simultaneously reduced in these countries. 
On the other hand, costs of PV and battery systems have seen precipitous falls in recent times.
These energy price hikes and asset cost reductions are driving customers to increase their levels of energy self-consumption by investing in energy storage technology, to complement rooftop PV systems. 

This presents a dilemma to distribution nework service providers (DNSPs) and vertically-integrated electricity utilities --- how to design tariffs that reflect the long-run marginal cost of electricity network assets, so that all consumers receive a price signal indicating the extent to which they each contribute to network peak demand, while (i) not encouraging customers with distributed energy resources (DER) to defect from the grid, and (ii) without unfairly apportioning network costs on customers without PV or other DER. 
This has proven to be a difficult task that has received much attention in the professional and academic literature~\cite{aemcrule,energy2014towards,lu2018designing,eutariff}. 

To this end, this paper proposes a probabilistic framework to enable DNSPs to test cost-reflectivity of various network tariffs, while considering various DER, including rooftop PV, battery storage and flexible loads. The framework integrates statistical models of PV generation, electricity demand and hot water use, a novel HEMS formulation that explicitly models peak demand charge, and a Monte Carlo (MC) power flow model to assess technical and economic impacts of network tariffs on distribution networks. 
This paper thus fills an important gap in the existing research, which so far considered either only technical or only economic aspects of the problem using deterministic tools.

In more detail, recent studies have considered economic impacts of \textit{energy-} and \textit{demand-based tariffs} on residential customers and on utilities' revenue. 
{Demand-based tariffs} have been shown to effectively resolve network price instability and reduce cross-subsidies between consumers without DER and prosumers~\cite{simshauser2016distribution}, and also to ensure a stable revenue for DNSPs~\cite{young2016electricity}. 
From the customer perspective,~\cite{abdelmotteleb2018designing} utilised a \textit{peak coincidence} network charge coupled with a fixed charge to reduce energy cost for price responsive customers. This slightly outperformed a peak demand charge but led to a reduction in overall system cost compared to traditional volumetric tariffs.

In~\cite{nijhuis2017analysis}, the authors suggested that a peak demand tariff based on a customer's yearly peak demand should be considered by DNSPs, as it performed best in terms of cost-reflectivity and predictability amongst other tariff types. 
On the contrary, {demand-based tariffs} proposed by the Australian Energy Regulator (AER) was tested on households in Sydney, from which it was concluded that without due adjustments made, these tariffs show low cost-reflectivity~\cite{passey2017designing}. 
From these studies, it is evident that the suitability of network tariffs in terms of cost-reflectivity is dependent on the assumptions made in the actual design and on how customers respond to these tariffs~\cite{stenner2015australian}.  

\par Despite these efforts, very little research has considered the technical impacts and consequences of network tariff designs on use of the distribution network. 
This is paramount because the aggregate network peak demand and energy losses are the long-run network cost drivers. 
It was shown in~\cite{pimm2018time} that \textit{time-of-use} (ToU) tariffs alone can increase peak loading on networks with deep DER penetration levels, where customers seek to maximise their cost savings. In view of this, authors in~\cite{supponen2016network} showed that {demand-based tariffs} could be used to mitigate transformer loading at medium voltage (MV) substations. 
Similarly, the results in~\cite{steen2016effects} demonstrated the effectiveness of {demand-based tariffs} in alleviating peak demand whilst considering demand response from customers' controllable appliances. 
In~\cite{steen2016effects}, however, customers were exposed to spot market prices (dynamic prices) and the effects of PV-battery systems were not considered.


\par Given this background, this paper extends our preliminary results~\cite{azuatalam2017impacts} to address two main problems:
\begin{itemize}
\item DNSPs currently don’t possess tools to assess the impact of network tariffs on peak demand, so we propose a probabilistic framework that supports the design of DER-specific cost-reflective tariffs;
\item Even when appropriate tariffs exist, they might not be effective due to the lack of customers’ response. Therefore, we also show how DER-specific tariffs can be complemented with an automated HEMS that allows customers to shift the demand while retaining the desired comfort level.
\end{itemize}

The proposed framework first generates synthetic traces of PV generation, electricity demand and electric hot water use, which are fed into a HEMS optimisation model
that determines the optimal DER schedule given the network tariff.
The HEMS optimisation is then run for 332 customers for a year to account for seasonal variations in demand and solar PV output.
Three scenarios are considered based on customer DER ownership, namely, electric water heater (EWH) only, EWH$+$PV, and EWH$+$PV$+$battery; and simulation is performed for four different network tariff types.
The output of the HEMS optimisation model, which determines the shape of the electric demand profile, is used in probabilistic power flow to examine the impact of the tariff types on typical low voltage (LV) distribution networks.


The HEMS optimisation model, based on mixed integer linear programming (MILP) has the objective of minimising customers' electricity cost under energy- and demand-based network tariffs, subject to device constraints and grid connection limits. For modelling {demand-based tariffs}, we include the peak demand charge as a linear term in the objective function corresponding to an additional peak demand variable multiplied by the set demand charge, which is incorporated into the model using an inequality constraint that sets the peak demand variable equal to the maximum monthly demand. In this way, we retain the computational efficiency of the MILP approach by avoiding the computationally expensive min-max formulation used in~\cite{steen2016effects} that models the peak demand explicitly. 
We build on our earlier work in~\cite{azuatalam2017impacts} by including electric water heaters as part of the HEMS formulation, since they account for a considerable portion of energy consumption in the Australian context and can affect peak loading~\cite{greenhouse2013}.

In summary, the proposed framework is underpinned by: 
(i) a novel home energy management formulation that explicitly considers peak demand charges while retaining the computational efficiency of the conventional MILP formulation; 
(ii) a principled statistical solar PV and demand model to synthesize a pool of residential load traces; and 
(iii) a principled statistical hot water use model to synthesize a pool of residential hot water use profiles.

To validate the methodology, we demonstrate the impacts of energy- and demand-based network tariffs on typical LV distribution networks. Specifically, we investigate the effects of these network tariffs on annual feeder head loading and customer voltage profiles at different PV-battery penetration levels.



%

The remainder of this paper is structured as follows: In the next section, we present an overview of the tariff assessment framework. Following this, in Section~\ref{synthetic}, we describe the steps to derive the solar PV/demand and EWH hot water use statistical models while Section~\ref{ham} outlines household DER modelling. Section~\ref{optimisation} details the optimisation model of the network tariff types and steps taken to perform annual electricity cost calculations. In Section~\ref{powerflow}, we describe the power flow analysis framework. The case study is described in Section~\ref{case_study} while the simulation results are presented and discussed in Section~\ref{results}. Section~\ref{conclusion} concludes the paper and suggests further work.


\section{Methodology overview} 

\begin{figure}[t]
	\centering
	\includegraphics[scale = 0.65]{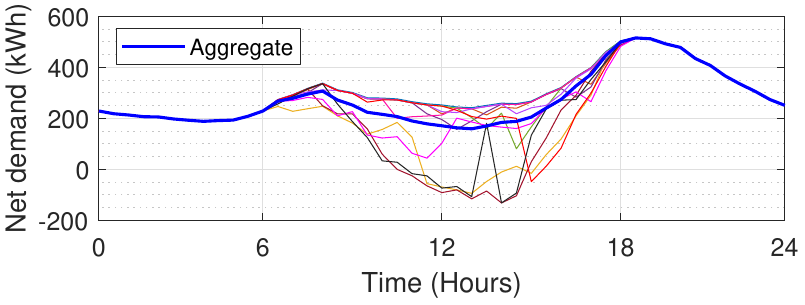}
	\caption{Weekday net demand profiles for a set of ten customers at 80\% PV penetration level and the aggregated net demand of the same ten customers.} 
	\label{net_demand}
\end{figure}

To evaluate the impact of network tariffs on customer response and the resultant effects on a LV network, 
it is imperative to model each customers' HEMS individually. 
As an illustration,
Fig.~\ref{net_demand} shows a set of ten individual net demand profiles at 80\% PV penetration level plotted against the aggregate net demand of the same ten customers. Observe that while the net demand of individual customers can be negative, implying power export to the grid, the aggregate profile is always positive; this goes to show that an aggregate demand model can be misleading.
In contrast, we model each customer individually. The statistically generated demand profiles are then randomly assigned to different locations in the network using a MC approach, which serves as an input for probabilistic load flow analysis.

A summary of the probabilistic assessment framework is detailed in Fig.~\ref{fig1}. In Module I, using yearly historical data, we generate a pool of net load traces and corresponding hot water use profiles by applying the statistical models of PV generation, electricity demand and hot water use, as described in Section~\ref{synthetic}.
In Module II, the outputs of the statistical models are fed as inputs to the MILP-based HEMS to solve the yearly optimisation problem for the different tariff types, and results are saved for each customer. 
The MILP-based HEMS is described in two parts--Sections~\ref{ham} and~\ref{optimisation}. 
First, Section~\ref{ham} describes detailed models of the battery energy storage system and the electric hot water system, which can be reused under different tariff designs and incentive structures.
Second, Section~\ref{optimisation} outlines the optimisation model, whose objective is to minimise customers' electricity cost under energy- and demand-based tariffs. Section~\ref{optimisation} also details the optimisation model for three scenarios based on DER ownership, and the cost implications of different tariff types. Based on this, the economic impacts of the network tariffs are analysed and discussed in Section~\ref{annual_cost}.

To assess the technical impacts of the network tariffs on the distribution network, we assume that the residential customers, with individually modelled HEMS and price response, all form part of a LV network. Hence, the optimisation results and output data from Module I are used to perform time-series yearly MC power flow studies on three representative LV distribution networks using OpenDSS~\cite{opendss} as described in Section~\ref{powerflow}. MC simulation is employed to cater for the uncertainties in customer location and the size of DER. Therefore, 100 MC power flow simulations are performed to investigate the impacts of the network tariff types on customer voltage profile and feeder head loading at different PV-battery penetration levels. 

\pgfdeclarelayer{background}
\pgfdeclarelayer{foreground}
\pgfsetlayers{background,main,foreground}

\tikzstyle{materia}=[draw, thick, text width=30em, text centered,
minimum height=1.5em, drop shadow]
\tikzstyle{practica} = [materia, fill=white, text width=28em, minimum width=28em,
minimum height=3em, rounded corners, drop shadow] 
\tikzstyle{texto} = [above, text width=25em]
\tikzstyle{linepart} = [draw, thick, -latex', dashed]
\tikzstyle{line} = [draw, thick, -latex']
\tikzstyle{ur}=[draw, text centered, minimum height=0.01em]

\newcommand{\blockdist}{1.3}
\newcommand{\edgedist}{1.5}

\newcommand{\practica}[2]{node (p#1) [practica]
	{\textbf{Step #1}\\{\normalsize\textit{#2}}}}

\newcommand{\background}[5]{%
	\begin{pgfonlayer}{background}
		\path (#1.west |- #2.north)+(-0.5,+0.5) node (a1) {};
		\path (#3.east |- #4.south)+(+0.5,-0.5) node (a2) {};
		\path[rounded corners, fill=lightgray!40, draw, thick, dashed] 
		(a1) rectangle (a2);
		\path (a1.east |- a1.south)+(4.6,-0.4) node (u1)[texto]
		{\normalsize\textit{Module #5}};
\end{pgfonlayer}}

\newcommand{\transreceptor}[3]{%
	\path [linepart] (#1.east) -- node [above]
	{\scriptsize Transreceptor #2} (#3);}

\begin{figure}[t!]
	\begin{center}
		
		\begin{tikzpicture}[scale=0.7,transform shape]
		
		\path \practica {1}{Generate a pool of net load traces \\ using the PV and Demand statistical model};
		\path (p1.south)+(0.0,-1.2) \practica{2}{Generate corresponding hot water use \\ profiles using the hot water use statistical model};

		\path (p2.south)+(0.0,-1.9) \practica{3}{Using tariff and DER data and output from Steps 1 and 2, solve the HEMS problem using MILP for a year};
		\path (p3.south)+(0.0,-1.2) \practica{4}{Save the yearly power import/export results and calculate the annual electricity cost for each customer};
		
		\path (p4.south)+(0.0,-1.9) \practica{5}{utilise data from Steps 1, 2 and the power exchange results from step 4 to run yearly Monte Carlo power flow};
		\path (p5.south)+(0.0,-1.2) \practica{6}{Save customer voltage profiles and feeder head loading for each MC simulation};

		\path [line] (p1.south) -- node [above] {} (p2);
		\path [line] (p2.south) -- node [above] {} (p3);
		\path [line] (p3.south) -- node [above] {} (p4);

		\path [line] (p1.east) -- +(0.34,0.0) -- +(0.34,-9.1) -- node [right] {} (p5);
		\path [line] (p2.west) -- +(-0.34,0) -- +(-0.34,-7.15) -- node [left] {} (p5);
		\path [line] (p4.south) -- node [above] {} (p5);

		\path [line] (p5.south) -- node [above] {} (p6);
		

		\background{p1}{p1}{p1}{p2}{I: Demand, PV and EWH Water Use Synthesis}
		\background{p3}{p3}{p3}{p4}{II: HEMS Problem}
		\background{p1}{p5}{p5}{p6}{III: MC Power Flow}
		
		\end{tikzpicture}
		\caption{Overview of the Methodology.}
		\label{fig1}
	\end{center}
\end{figure}
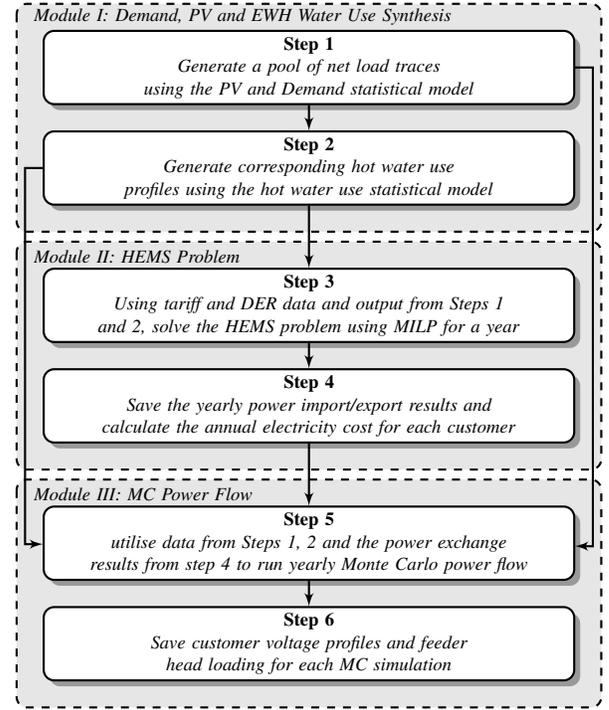

\section{Demand, solar PV and electric hot water use statistical models} \label{synthetic}
In order to perform a probabilistic assessment of the impact of flexible loads in LV distribution networks under various network tariffs, a large pool of PV, demand and EWH profiles are required. To this end, we provide models to generate representative profiles using principled statistical approaches.

\begin{figure}[t]
	\centering
	\hbox{\hspace{0.65em}\includegraphics[scale = 0.65]{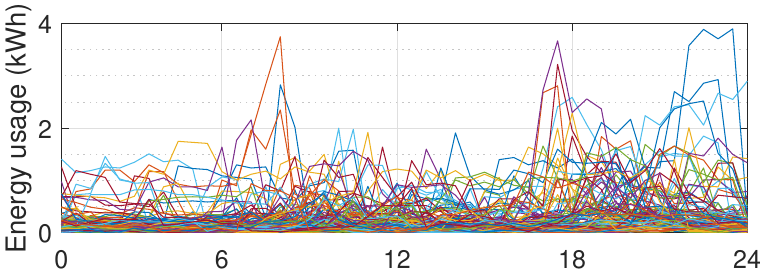}}
	\hbox{\hspace{-0.2em}\includegraphics[scale = 0.65]{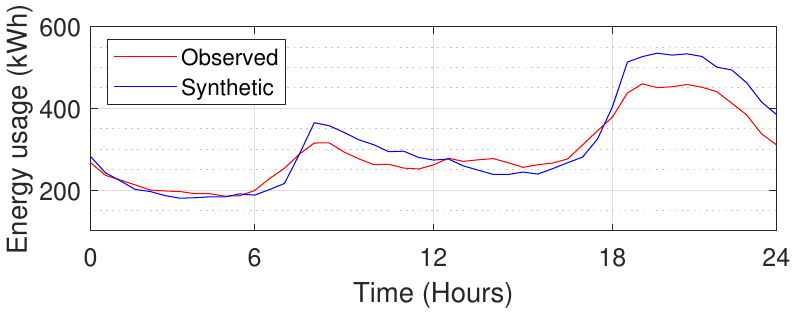}}
	\caption{1000 synthetic demand profiles (top); aggregated observed and synthetic weekday demand profiles (bottom).}
	\label{synthetic_profiles}
\end{figure}

\subsection{Demand and Solar PV Statistical Models} \label{pv-demand}
In this section, we extend the nonparametric Bayesian model introduced in~\cite{power2017nonparametric} to generate a pool of demand and PV profiles needed to perform probabilistic power flow studies. To accomplish this, we first cluster historical data sourced from the Ausgrid \textit{Solar Home Electricity Data} into representative clusters, using the MAP-DP (\textit{maximum a-posteriori Dirichlet process mixtures}) technique. Next, we employ the Bayesian estimation method to estimate the probability that an unobserved customer possesses certain features identified in particular clusters.
The number of occurrence of these features (count) is used as a hyperparameter of a \textit{Dirichlet distribution} $\mathop{\mathrm{Dir(\alpha)}}$.
 
%
%

To assign a cluster to an unobserved customer, we use a random variable drawn from a \textit{categorical distribution} $\mathop{\mathrm{Cat(\gamma)}} $ over the features of the particular cluster, where the parameters $\gamma$ are obtained by sampling from $\mathop{\mathrm{Dir(\alpha)}}$. We then generate a pool of net load traces specific to assigned features based on a Markov chain process.  More details on the PV and demand statistical model can be found in~\cite{2018arXiv180800615P}.

The solar PV and demand statistical models were cross-validated in~\cite{2018arXiv180800615P}, using the \textit{Smart Grid Smart City} data set. 
As an illustration, consider the comparison between 1000 synthetic demand profiles (Fig.~\ref{synthetic_profiles}, top) and the aggregated demand profile of the 1000 customers plotted against the observed demand profile that was used to generate the synthetic data (Fig.~\ref{synthetic_profiles}, bottom). You can observe a very good match, with the mean absolute error of 9.80\% in this case.

\subsection{Electric Hot Water Use Statistical Model} \label{hwd}
The hot water statistical model is defined for aggregated intervals of time slots during the day. It comprises a location distribution within an interval and a magnitude distribution for each time slot. The model is estimated following three steps. 
First the data is broken into \textit{intervals} of the day, comprised of sets of contiguous time slots.
The specific intervals used in this work are given in Table~\ref{table8}.

Second, a \textit{location process} is estimated for each interval. This consists of a distribution over the number of draws in an interval, and is given by a homogeneous \textit{Poisson distribution}, $\mathop{\mathrm{Poi(\mu)}}$, given by:
\begin{equation} \label{eqn11}
P(k\ \mathrm{draws\ in\ interval})= 
\exp\left[-\mu \right] \frac{\mu^{k}}{k!}
\end{equation} where $\mu>0$ is the rate of draw events during the interval.

Third, a magnitude distribution is estimated for the size of the draws in each interval. The magnitude of the draws are modeled as following a \textit{Weibull distribution} $\mathop{\mathrm{Wei(\kappa,\sigma)}}$, given by:


\begin{equation}
f(x | \kappa,\sigma) = \twopartdef { \frac{\sigma}{\kappa}\left(\frac{x}{\kappa}\right)^{\sigma-1} 
\exp\left[ -\left(\frac{x}{\kappa} \right)^\sigma \right] } 
{x \geq 0} {0} {x < 0}
\end{equation} where $\kappa>0$ is a \textit{scale} parameter and $\sigma>0$ is a \textit{shape} parameter.



Sampling from this model involves one additional element. 
Specifically, once the models are estimated and values of $\mu$, $\kappa$ and $\sigma$ computed, the full sampling process for an interval involves: 	(i) sampling a number of draws in an interval according to $\mathop{\mathrm{Poi(\mu)}}$ (ii) allocating these draws to time slots over the interval's time slots according to a \textit{uniform distribution} and (iii) sampling draw sizes for each draw according to $\mathop{\mathrm{Wei(\kappa,\sigma)}}$. We emphasize that in order to sample time slots for hot water draws, each interval first has a number of draws sampled from the estimated \textit{Poisson distribution}, and then that number of locations are allocated to draws in the interval according to a uniform distribution (with replacement) over time slots, as is the standard approach for sampling from Poisson processes.
Unlike for the demand and PV traces, cross-validation for the EWH traces was not possible due to the lack of empirical hot water time-of-use data.

%

%
%

\begin{table}[t]
	\footnotesize
	\centering
	\caption{HW model intervals, with time slots indicated by their start time.}
	\label{table8}
	\begin{tabular}{cc|cc}
		\hline
		Begin & End & Begin & End \\
		\hline
		23:00 & 1:30 & 11:00 & 13:30 \\
		2:00 & 4:30 & 14:00 & 16:30 \\ 
		5:00 & 7:30 & 17:00 & 19:30 \\ 
		8:00 & 10:30 & 20:00 & 22:30 \\ 
		\hline
	\end{tabular}

\end{table}

\section{Household DER modelling} \label{ham}

For each customer, $c \in \mathcal{C}$ possessing a set of appliances, $\mathcal{A}:= \{1,2,...,\left\vert\mathcal{A}\right\vert\}$, let $\alpha \in \{1,...,M\}$ denote customer's $c$ appliance type, wherefore $\mathcal{A}_\alpha \subseteq \mathcal{A}$. In this work, we consider just three (3) appliance types $(M = 3)$: Type 1 set includes energy storage devices, particularly batteries; 
Type 2 set includes thermostatically-controlled devices, particularly electric water heaters (EWH);
Type 3 appliances constitute the base load and includes all must-run and uncontrollable devices.

%
%


%
%
%

\subsection{Battery Energy Storage System (BESS) Modelling}
The BESS operational model is linearised so that it fits the MILP optimisation framework. Battery sizes utilised in this study range from 6 to 12~\si{kWh} and are obtained from ZEN Energy~\cite{zenenergy}. We have assumed a minimum/maximum battery SOC of 10\%/100\% nominal capacity and a round-trip efficiency of 90\% for all battery sizes. For all $a \in \mathcal{A}_1, h \in \mathcal{H}$: 

\begin{align}
& e_{a,h}^\mathrm{b} = e_{a,h-1}^\mathrm{b} + \Delta h \Big(\eta_a^\mathrm{b+}p_{a,h-1}^\mathrm{b+} -\ \big(1/\eta_a^\mathrm{b-}\big)p_{a,h-1}^\mathrm{b-} \Big)  \label{eqn1} \\
& p_{a,h}^\mathrm{b+}\; \leq\; \bar{p}^\mathrm{b+}s_{a,h}^\mathrm{b}  \label{eqn2}\\ 
& p_{a,h}^\mathrm{b-}\; \leq\; \bar{p}^\mathrm{b-}\big(1 - s_{a,h}^\mathrm{b}\big)  \label{eqn3}\\  
& 0\; \leq\; p_{a,h}^\mathrm{b+} \leq\; \bar{p}^\mathrm{b+} \label{eqn4}\\ 
&  0\; \leq\; p_{a,h}^\mathrm{b-} \leq\; \bar{p}^\mathrm{b-} \label{eqn5}\\ 
& \barbelow{e}^\mathrm{b}\; \leq\; e_{a,h}^\mathrm{b} \leq\; \bar{e}^\mathrm{b} \label{eqn6}
\end{align}


\begin{table}[t]
	\footnotesize
	\centering
	\caption{EWH Parameters}
	\label{table6}
	\begin{tabular}{cccc}
		\hline 
		Number of & EWH & Element & Tank surface \\
		Customers &  Size $(V)$ & rating $(Q)$  & Area $(A)$ \\
		\si{\percent} & \si{Liter} & \si{kW} & \si{\meter \squared} \\
		\hline
		2.44 & 80 & 1.8 & 1.114 \\ 
		8.94 & 125 & 3.6 & 1.500 \\
		86.99 & 160 & 3.6 & 1.768 \\
		1.63 & 250 & 4.8 & 2.381 \\
		\hline 
		Density $(\rho)$ & Specific heat $(c)$ & $T_\mathrm{in}$ range & Conductance $(U)$  \\
		\si{\kilogram/\meter \cubed}  & \si{\kilo \joule/\kilogram. \degreeCelsius} & \si{\degreeCelsius} & \si{\watt/\meter \squared. \degreeCelsius} \\ \hline
		1000  & 4.18 & 60 - 82 & 1.00 \\
		\hline
	\end{tabular}
\end{table}


\subsection{Electric Water Heater (EWH) Modelling}

The EWH operational model is given by a set of difference equations in order to fit them into an optimisation model~\cite{kar1996optimum,elamari2011using}. We consider single-element EWH tanks from Rheem\footnote{Rheem Electric Storage Water Heaters Specification Sheet http://www.rheem.com.au/DomesticElectricWaterHeaters} and estimated the EWH sizes for the 123 selected customers using their hot water profiles. The EWH simulation parameters are given in Table~\ref{table6}. For all $a \in \mathcal{A}_2, h \in \mathcal{H}$:
\begin{align}
& p_{a,h} = \eta_a^\mathrm{th} u_{a,h}^\mathrm{th}Q_a  \label{eqn7}	 \\
&\begin{aligned} 
& T_{a,h}^\mathrm{in} = T_{a,h-1}^\mathrm{in} + \psi_a p_{a,h} + \lambda_a(T_{a,h-1}^\mathrm{out} - T_{a,h-1}^\mathrm{in})\  \\                                                                         & \hspace{10em} + \phi_a(T_{a,h-1}^\mathrm{inlet} - T_{a,h-1}^\mathrm{in}) 
\end{aligned} \label{eqn8} \\
& T_{a,h}^\mathrm{in,min} \leq  T_{a,h}^\mathrm{in} \leq T_{a,h}^\mathrm{in,max} \label{eqn9}
\end{align} where: $C = \rho V c$; $A\approx 6V^{2/3}$;
$ \psi_a = \cfrac{\Delta h}{C}$;  $\lambda_a = \cfrac{UA \Delta h}{C}$; $\phi_a = \rho W_d$;  $W_d$ = EWH water use in liters; $Q_a$ = EWH element rating in \si{kW}.

\par The second term at the RHS of~\eqref{eqn8} represents the energy from the resistive element of the EWH. The third term represents the heat losses to the ambient, while the last term represents the energy required to heat the inlet cold water.

\section{Optimisation model and electricity cost calculations} \label{optimisation}
In this section, the optimisation model for all tariff types considering customers with EWH and PV-battery installed is described. Each problem is solved for a year, using a rolling horizon approach and a monthly decision horizon. For customers with just EWH and solar PV, the models are modified accordingly by removing the battery parameters as described in Section~\ref{optscen}. In Section~\ref{annualcost}, we provide the formulas for computing annual electricity cost for each tariff type.


\subsection{Optimisation Model for Energy-based Tariffs}
For customers facing an \textit{energy-based tariff} (\textit{Flat} or \textit{ToU}) the monthly optimisation model is given in~\eqref{eqn12} to~\eqref{eqn19} for all $h \in \mathcal{H}$:



\begin{align} \label{eqn12}	
\underset{\substack{p_{d',h}^\mathrm{g+}, p_{d',h}^\mathrm{g-}, p_{d',h}^\mathrm{b+},\\ p_{d',h}^\mathrm{b-}, p_{d',h}^\mathrm{d},  d_{d',h}^\mathrm{g}, \\s_{d',h}^\mathrm{b}, e_{d',h}^\mathrm{b}, u_{d',h}^\mathrm{th}, \\T_{d',h}^\mathrm{in}}}{\text{minimise}}
& \ \sum\limits_{d' \in \mathcal{D}'}  \bigg[ \sum\limits_{h \in \mathcal{H}}  T^\mathrm{flt/tou}  p_{d',h}^\mathrm{g+} - T^\mathrm{fit}p_{d',h}^\mathrm{g-}  \bigg]  \\
\text{subject to} \ \ & \text{\cref{eqn1,eqn2,eqn3,eqn4,eqn5,eqn6,eqn7,eqn8,eqn9}} \quad \nonumber \\
&\begin{aligned}
\hspace{-2cm}	& p_{d',h}^\mathrm{g+} - p_{d',h}^\mathrm{g-} = \eta^\mathrm{i} \Big( p_{d',h}^\mathrm{b+} - p_{d',h}^\mathrm{b-} - p_{d',h}^\mathrm{pv} \Big) + p_{d',h}^\mathrm{d} \\
\end{aligned} \label{eqn14} \\
&\begin{aligned}
\hspace{-2cm}	& p_{d',h}^\mathrm{d} = p_{h}^\mathrm{base} + \sum\limits_{a \in \mathcal{A}_2} p_{a,d',h} \\
\end{aligned} \label{eqn15} \\
& \hspace{-1.9cm} p_{d',h}^\mathrm{g+}\; \leq\; \bar{p}^\mathrm{g}d_{d',h}^\mathrm{g} \label{eqn16} \\ 
& \hspace{-1.9cm}  p_{d',h}^\mathrm{g-}\; \leq\; \bar{p}^\mathrm{g}\big(1 - d_{d',h}^\mathrm{g}\big) \label{eqn17} \\  
& \hspace{-1.9cm}  0\; \leq\; p_{d',h}^\mathrm{g+} \leq\; \bar{p}^\mathrm{g} \label{eqn18} \\ \hspace{3em} 
& \hspace{-1.9cm}  0\; \leq\; p_{d',h}^\mathrm{g-} \leq\; \bar{p}^\mathrm{g} \label{eqn19} 
\end{align}

\subsection{Optimisation Model for Demand-based Tariffs}
 

For customers facing a \textit{demand-based tariff} (\textit{FlatD} or \textit{ToUD}), an additional constraint~\eqref{eqn22} is used to limit the grid import according to the demand charge component, $T^\mathrm{pk}\hat{p}$ in~\eqref{eqn20}. This does not explicitly model demand charge as in practice, but implicitly achieves the same objective of clipping a customer's peak demand (See Fig.~\ref{daily_peak}). The monthly optimisation model is given below for all $h \in \mathcal{H}$:

\begin{align} \label{eqn20}	
\underset{\substack{p_{d',h}^\mathrm{g+}, p_{d',h}^\mathrm{g-}, p_{d',h}^\mathrm{b+},\\ p_{d',h}^\mathrm{b-}, p_{d',h}^\mathrm{d},  d_{d',h}^\mathrm{g}, \\ s_{d',h}^\mathrm{b}, e_{d',h}^\mathrm{b}, u_{h}^\mathrm{th}, \\ T_{d',h}^\mathrm{in}, \hat{p}}}{\text{minimise}}
&  T^\mathrm{pk}\hat{p} + \sum\limits_{d' \in \mathcal{D}'}  \bigg[ \sum\limits_{h \in \mathcal{H}}  T^\mathrm{flt/tou} p_{d',h}^\mathrm{g+} - T^\mathrm{fit}p_{d',h}^\mathrm{g-} \bigg] \\
\text{subject to} \ \ & \text{\cref{eqn1,eqn2,eqn3,eqn4,eqn5,eqn6,eqn7,eqn8,eqn9,eqn14,eqn15,eqn16,eqn17,eqn18,eqn19}} \quad \nonumber\\
& \hspace{-1.9cm} p_{d',h}^\mathrm{g+} \leq  \hat{p}  \label{eqn22}  
\end{align}

\subsection{Optimisation Scenarios} \label{optscen}
\par The optimisation models described above are solved for three scenarios based on customer DER ownership. Scenario I is the base case where all customers possess just EWH. Then we progressively add DER to form the other two scenarios, following~\eqref{eqn14}. Where $p_{h}^\mathrm{d} = p_{h}^\mathrm{base} + p_{h}^\mathrm{ewh}$, then the following scenarios hold:

\begin{itemize}
    \item Scenario I: The energy balance equation for customers with EWH only is:
\begin{equation}
p_{h}^\mathrm{g+} = p_{h}^\mathrm{d}  \label{eqn23}
\end{equation}

\item Scenario II: The energy balance equation for customers with EWH and solar PV is:
\begin{equation}
p_{h}^\mathrm{g+} - p_{h}^\mathrm{g-} = -\eta^\mathrm{i} p_{h}^\mathrm{pv}  + p_{h}^\mathrm{d} \label{eqn24}
\end{equation}

\item Scenario III: The energy balance equation for customers with EWH, solar PV and batteries is:
\begin{equation} 
p_{h}^\mathrm{g+} - p_{h}^\mathrm{g-} = \eta^\mathrm{i} \Big(p_{h}^\mathrm{b+} - p_{h}^\mathrm{b-} - p_{h}^\mathrm{pv} \Big) + p_{h}^\mathrm{d} \label{eqn25}
\end{equation}

\end{itemize}

\subsection{Annual Electricity Cost Calculations} \label{annualcost}
The annual electricity cost for customers with PV or PV-battery (Scenarios II and III) are calculated for each tariff type as in~\eqref{eqn26} to~\eqref{eqn29} using $p_{d',h}^\mathrm{g+}$ and $p_{d',h}^\mathrm{g-}$, obtained as output variables from the optimisation. For customers without DER (Scen. I), the calculations are done without the power export component ($T^\mathrm{fit}p_{d',h}^\mathrm{g-}$). 


\begin{equation} \label{eqn26}
\mathrm{C(\textit{Flat})} = \sum\limits_{d \in \mathcal{D}} \bigg[ T_{d}^\mathrm{fx} +  \sum\limits_{h \in \mathcal{H}} \Big( T^\mathrm{flt}p_{d,h}^\mathrm{g+} -  T^\mathrm{fit}p_{d,h}^\mathrm{g-}  \Big)\Delta h \bigg]
\end{equation}

\begin{equation} \label{eqn27}
\mathrm{C(\textit{ToU})} = \sum\limits_{d \in \mathcal{D}} \bigg[ T_{d}^\mathrm{fx} +  \sum\limits_{h \in \mathcal{H}} \Big( T_{h}^\mathrm{tou}p_{d,h}^\mathrm{g+} -  T^\mathrm{fit}p_{d,h}^\mathrm{g-}  \Big)\Delta h \bigg]
\end{equation}

\begin{multline} \label{eqn28}
\mathrm{C(\textit{FlatD})} = \sum\limits_{d \in \mathcal{D}} \bigg[ T_{d}^\mathrm{fx} + \sum\limits_{h \in \mathcal{H}} \Big( T^\mathrm{flt}p_{d,h}^\mathrm{g+} -  T^\mathrm{fit}p_{d,h}^\mathrm{g-}  \Big)\Delta h \bigg] \\ + \sum\limits_{m \in \mathcal{M}} \Big( T^\mathrm{pk}p_{m}^\mathrm{pk} \Big)
\end{multline}

\begin{multline} \label{eqn29}
\mathrm{C(\textit{ToUD})} = \sum\limits_{d \in \mathcal{D}} \bigg[ T_{d}^\mathrm{fx} + \sum\limits_{h \in \mathcal{H}} \Big( T_{h}^\mathrm{tou}p_{d,h}^\mathrm{g+} -  T^\mathrm{fit}p_{d,h}^\mathrm{g-}  \Big)\Delta h  \bigg] \\ 
+ \sum\limits_{m \in \mathcal{M}} \Big( T^\mathrm{pk}p_{m}^\mathrm{pk} \Big)
\end{multline}


\par The value $p_{m}^\mathrm{pk}$ is calculated either based on the peak monthly demand (\textit{FlatD} and \textit{ToUD}) or on the average top four daily peak demand (\textit{FlatD4} and \textit{ToUD4}) for each month. In essence,  the \textit{demand-based tariffs} each has two variants based on the calculation of the monthly peak demand.

\section{Power flow analysis} \label{powerflow}

\par We consider a LV distribution network as a radial system denoted $\mathcal{G(N,E)}$. This comprises of $\left\vert \mathcal{N}\right\vert$ nodes in the set $\mathcal{N}:= \{0,1,...,N\}$ representing network buses, and distribution lines, each denoted as a tuple $(i,j)$ connecting the nodes and represented by the set of edges $\mathcal{E}:= \{(i,j)\} \subset \mathcal{\{N \times N\}}$. Each customer, $c \in \mathcal{C}$ in the network is connected to a load bus as a single-phase load point, where the number of load buses $\left\vert \mathcal{N}_c\right\vert$ is a subset of the total nodes in the network (and $\mathcal{N}_c \subseteq \mathcal{N}$). Let $\boldsymbol{V} = [v_0, v_1,...,v_N]$ be the voltage magnitudes at the nodes, where $v_0$ is the substation voltage. Let $v_c$ be the voltage at each (customer) load point. These voltages are monitored at every half-hour in the year to check for any voltage violations. More so, the current flowing through the line connecting nodes 0 and 1 (denoted $i^\mathrm{head}$) is monitored to check for any thermal loading problems. We assume that each customer, $c \in \mathcal{C}$ in the network utilises a HEMS to manage a set of appliances in order to minimise electricity cost.

\par The net grid power exchange ($p_{d}^\mathrm{g} = p_{d}^\mathrm{g+} - p_{d}^\mathrm{g-}$) resulting from the HEMS optimisation solution and the data generated from the statistical models (see Module III, Step 5 in Fig.~\ref{fig1}) are fed as input to a distribution network model to perform MC power flow analysis, using Algorithm~\ref{MCPF Algorithm}. We then carry out a probabilistic assessment of yearly voltage profiles ($v_{d,c}$) for each customer and feeder head loading ($i^\mathrm{head}_{d}$) in order to ascertain the level of voltage and thermal loading problems associated with any particular network. The definitions of voltage and thermal loading problem are:


\begin{algorithm}[t]
	\caption{Monte Carlo power flow algorithm}\label{MCPF Algorithm}
	\footnotesize
	
	$\mathcal{P}$: set of PV penetration levels, $\mathcal{P}:= \{0,25,50,75\}$\\
	$\mathcal{B}$: set of battery penetration levels, $\mathcal{B}:= \{0,40,80\}$\\
		$\mathcal{C}$: set of customers in a LV network, $\mathcal{C}:= \{1,2,...,\left\vert\mathcal{C}\right\vert\}$ \\

	\begin{algorithmic}[1]
		\For{each $p \in \mathcal{P}$}
		\State Read yearly load and PV profile

		\If{$p = 0$}
		  \State \textbf{Read} $p_{d,c}^\mathrm{g}\ \ \forall\ c \in \mathcal{C}, d \in \mathcal{D}$, for Sc.I\Comment{base case: 0\% PV-battery}
		  \For{$k \longleftarrow 1\ \mathbf{to}\ 100\ \mathbf{step}\ 1$}\Comment{100 MC simulations}
		  \State \textbf{Sample} uniformly from $p_{d,c}^\mathrm{g,Sc.I}$ for allocation to load points. 
		  \State \textbf{Run} yearly power flow 
		  \State \textbf{Return} $i_d^{\mathrm{head},k}$ and $v_{d,c}^{k}$, $\forall\ c \in \mathcal{C}, d \in \mathcal{D}$
		  
		  \EndFor
		
		\Else
		
		   \For{each $b \in \mathcal{B}$}
		   
		      \State \textbf{Read} $p_{d,c}^\mathrm{g}\ \ \forall\ c \in \mathcal{C}, d \in \mathcal{D}$, for Sc. I, II and III.
		   \For{$k \longleftarrow 1\ \mathbf{to}\ 100\ \mathbf{step}\ 1$}\Comment{100 MC simulations}
	    	\State $p_{d,c}^\mathrm{g, Sc.I}$:=  $(100 - p)\%$ of $p_{d,c}^\mathrm{g, Sc.I}$ + $p\%$.$(100 - b)\%$ of  \\\hspace{1.6cm} $p_{d,c}^\mathrm{g, Sc.II}$ + $p\%$.$b\%$ of $p_{d,c}^\mathrm{g, Sc.III}$ 
		    \State \textbf{Repeat} Lines 6 to 8
		   \EndFor
		
		   \EndFor
		
		\EndIf

		\EndFor
	\end{algorithmic}
\end{algorithm}


	\begin{itemize}
\item If a customer's voltage goes outside the range $0.95\ \mathrm{pu} \leq v_{d,c} \leq 1.05\ \mathrm{pu}$ during \SI{95}{\%} of days in a year, the customer is said to have a voltage problem \cite{essentialstd}. 
\item If the current flowing through line $i^\mathrm{head}_{d}$ (feeder head) exceeds its thermal rating, there is a thermal loading problem in the network. 
	\end{itemize}

\section{Case Study} \label{case_study}
Here, we provide the necessary data for our case study. This includes network data for three representative LV networks, the network tariff and retail charges and the customer demand and DER data.

\subsection{Low Voltage Networks}
\par The low voltage network data used in this work were obtained from the \textit{Low Voltage Network Solutions Project}~\cite{enwlreport}. Table~\ref{table1} summarizes the main features of the three networks used as case studies in this work. These are residential LV networks of different lengths and number of load points: Feeders 1 and 2 are fairly balanced while Feeder 3 is unbalanced. Given that these feeders are from the UK, we have modified them to suit the Australian context. Typical Australian LV networks are more robust with higher load capacity when compared to that from the UK. Therefore, we have increased the transformer capacity by a factor of three and decreased the line impedances by a factor of three since the average consumption in Australia is roughly three times that in the UK. However, the overall structure of LV networks in both countries are similar.

\begin{table}[t]
	\footnotesize
	\centering
	\caption{Network data}
	\label{table1}
	\begin{tabular}{cccc}
	\hline
		Feeder & Number of  & Total length & Feeder head\\
		number & customers & of all lines (\si{\meter}) & ampacity (\si{\ampere})\\
		\hline
		1 &  175 & 5206 & 1200 \\
		2 & 186 & 4197 & 1200\\
		3 & 302 & 10235 & 1155 \\
		\hline
	\end{tabular}
\end{table}

\subsection{Network Tariffs and Retail Charges}
\par A typical residential customer retail bill consists of network (distribution and transmission) charges, generation costs for energy, retailer's charge and other related costs. We have sourced the network tariff data, shown in Table~\ref{table2}), from Essential Energy\footnote{Essential Energy Network Price List and Explanatory Notes. Available at https://www.essentialenergy.com.au}. These are assumed fixed and known in advance. The peak demand charge in \si{\$/kW/month} is the charge for a customer's monthly peak demand (or, alternatively, the the average of the top four daily peak demand of a customer in a month). In Table~\ref{table3}, the residential electricity prices for customers in the Essential energy distribution zone for retailer, Origin Energy\footnote{Origin Energy NSW Residential Energy Price Fact Sheet for Essential Energy Distribution Zone. Available at https://www.originenergy.com.au}, is shown. These prices comprise the actual cost of electricity, retailer's service fee, and the network charge. In this study, we have assumed that the retailers pass on the DNSP tariff structure to the consumers. The different network tariffs (energy, \textit{Flat} and \textit{ToU}, and demand-based, \textit{FlatD} and \textit{ToUD}) are described below:


\begin{itemize}
	\item \textit{LV Residential Anytime (Flat)}: Includes a fixed daily charge and a flat usage charge.
	\item \textit{LV Residential Time-of-use (ToU)\footnote{Peak period:  7am to 9am, 5pm to 8pm; shoulder period: 9am to 5pm, 8pm to 10pm; off-peak period: 10pm to 7am.}}: Includes a fixed daily charge and a ToU usage charge.
	\item \textit{Small Residential - Opt in Demand Anytime (FlatD)}: Includes a fixed daily charge, a flat usage charge and a peak demand charge. 
	\item \textit{Small Residential - Opt in Demand (ToUD)}: Includes a fixed daily charge, a ToU usage charge and a peak demand charge.
\end{itemize}

\begin{table}[t]
	\footnotesize
	\centering
	\caption{Network tariff data}
	\label{table2}
	\begin{tabular}[t]{c@{\hspace{0.2cm}}c@{\hspace{0.2cm}}c@{\hspace{0.15cm}}c@{\hspace{0.15cm}}c@{\hspace{0.15cm}}c@{\hspace{0.15cm}}c@{\hspace{0.1cm}}}
		\hline 
		\multicolumn{1}{c}{\begin{tabular}[c]{@{}c@{}}Tariff\\ Type\end{tabular}} & \begin{tabular}[c]{@{}c@{}}Fixed\\ charge\\ \si{\$/day}\end{tabular} & \begin{tabular}[c]{@{}c@{}}Anytime\\ Energy\\ \si{c/kWh}\end{tabular} & \begin{tabular}[c]{@{}c@{}}Off peak\\ Energy\\ \si{c/kWh}\end{tabular} & \begin{tabular}[c]{@{}c@{}}Shoulder\\ Energy\\ \si{c/kWh}\end{tabular} & \begin{tabular}[c]{@{}c@{}}Peak\\ Energy\\ \si{c/kWh}\end{tabular} & \begin{tabular}[c]{@{}c@{}}Demand\\ Charge\\ \si{\$/kW/month}\end{tabular} \\ \hline
		
		
		\textit{Flat} & 0.8568 & 11.0321 & - & - & - & - \\ 
		\textit{ToU} & 0.8568 & - & 4.6287 & 12.6922 & 13.9934 & - \\ 
		\textit{FlatD} & 0.8568 & 3.2169 & - & - & - & 4.2112 \\ 
		\textit{ToUD} & 0.8568 & - & 2.1419 & 3.4771 & 4.0804 & 4.2112 \\ \hline
	\end{tabular}
\end{table}

\begin{table}[t]
	\footnotesize
	\centering
	\caption{Retail tariff data}
	\label{table3}
	\begin{tabular}[t]{c@{\hspace{0.25cm}}c@{\hspace{0.25cm}}c@{\hspace{0.25cm}}c@{\hspace{0.25cm}}c@{\hspace{0.25cm}}c@{\hspace{0.2cm}}c@{\hspace{0.1cm}}}
		\hline 
		\multicolumn{1}{c}{\begin{tabular}[c]{@{}c@{}}Tariff\\ Type\end{tabular}} & \begin{tabular}[c]{@{}c@{}}Fixed\\ charge\\ \si{\$/day}\end{tabular} & \begin{tabular}[c]{@{}c@{}}Anytime\\ Energy\\ \si{c/kWh}\end{tabular} & \begin{tabular}[c]{@{}c@{}}Off peak\\ Energy\\ \si{c/kWh}\end{tabular} & \begin{tabular}[c]{@{}c@{}}Shoulder\\ Energy\\ \si{c/kWh}\end{tabular} & \begin{tabular}[c]{@{}c@{}}Peak\\ Energy\\ \si{c/kWh}\end{tabular} & \begin{tabular}[c]{@{}c@{}}Feed-in\\ Tariff\\ \si{c/kWh}\end{tabular} \\ \hline
		
		
		\textit{Flat} & 1.5511 & 31.3170 & - & - & - & 9.0 \\ 
		\textit{ToU} & 1.5511 & - & 21.3400 & 37.1470 & 38.5880 & 9.0 \\ 
		\textit{FlatD} & 1.5511 & 23.5018 & - & - & - & 9.0 \\ 
		\textit{ToUD} & 1.5511 & - & 18.8532 & 27.9319 & 28.6750 & 9.0 \\ \hline
	\end{tabular}
\end{table}

\subsection{Customer Demand and DER Data}

\par We sourced the demand and solar PV generation data from the Ausgrid (DNSP in NSW) \textit{Solar Home Electricity Data}~\cite{solarhome}. This dataset comprises three years of half-hourly resolution smart meter data for the period between July 2010 to June 2013, for 300 residential customers in the Sydney region of Australia. The most recent data (for financial year, July 2012 to June 2013) is used in this study because it is complete and of higher quality, compared to the previous years in the dataset. Given that the \textit{Solar Home Electricity Data} do not contain customer hot water usage data, we selected 123 customers from the Ausgrid \textit{Smart Grid, Smart City} (SGSC)~\cite{smartgrid} dataset with complete hot water usage, solar PV and uncontrolled demand data. Then we randomly allocated these hot water profiles to selected 123 customers from the \textit{Solar Home Electricity Data}. 

\par Since the average PV size of the customers in the \textit{Solar Home Electricity Data} is roughly 1.5~\si{kW}, we applied a heuristic to update the PV sizes to reflect the current PV uptake rates and the average size of installed PV systems in Australia. The updated average PV size of these customers is roughly 4~\si{kW} and sizes range from 3 to 10~\si{kWp}, depending on the needs of the household. For customers with solar PV and batteries installed, the battery size of the customer depends on the size of the solar PV installed. In Australia, typically, 1.5-3~\si{kWh} of storage is used per 1~\si{kW} of PV installed~\cite{aemosmall}. This assumption is made in this work. The PV inverter efficiency has already been accounted for in the dataset, so we have assumed a PV inverter efficiency of 1 in our simulations. Table~\ref{table4} shows the PV-battery size combinations for the selected 123 customers with updated PV sizes.

\begin{table}[t]
	\footnotesize
	\centering
	\caption{PV-Battery size combinations}
	\label{table4}
	\begin{tabular}{ccc}
		\hline 
		Customers & Solar PV size & Battery size \\
		\si{\percent} & \si{kW} & \si{kWh} \\
		\hline
		76.42 & 3 - 4 & 6 \\
		20.33 & 5 - 6 & 8 \\
		2.44  & 7 - 8 & 10 \\
		0.81  & 9 - 10 & 12 \\
		\hline
	\end{tabular}
\end{table}

\section{Results and Discussion} \label{results}
In this section, the results from the optimisation and network power flows are analysed and discussed. First, we show the economic implications of the various network tariffs by carrying out annual electricity cost calculations (Section~\ref{annual_cost}). For this, 332 customers have been chosen from the generated pool of customers, since the largest feeder used as case study comprises 302 customers. Following this, the impact of network tariffs on customer daily and monthly peak demand is discussed (Section~\ref{daily_monthly}). Finally, the technical impacts on the network, of the different tariffs, are analysed in Sections~\ref{line_loading} and \ref{voltage_profile}.

\subsection{Annual Electricity Cost} \label{annual_cost}

\par
\begin{figure}[t]
	\centering
	\includegraphics[scale=0.65]{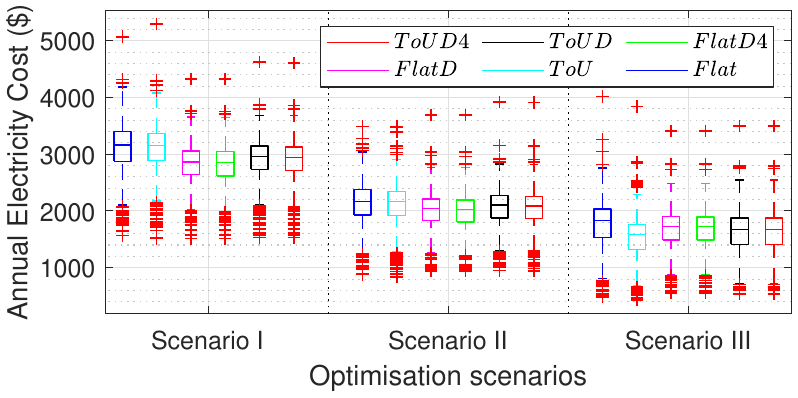}
	\caption{Annual electricity cost for 332 in the three scenarios.}
	\label{annualcosts}
\end{figure}

 In this section, we analyse the annual electricity costs for all scenarios using the results from Section~\ref{annualcost}, as illustrated in Fig.~\ref{annualcosts}. Overall, customers pay less for electricity as DER is progressively added. While \textit{demand-based tariffs} result in a lower electricity cost compared to \textit{energy-based tariffs} in Scenario I, this slightly levels off in Scenarios II and III. This is because when prosumers' grid power import is clipped due to demand charges, they compensate for this by exporting more power to the grid (via FiT payments). Nevertheless, the FiT rates are small compared to the retail rates so the net savings are minimal. With PV and batteries (Scenario III), however, large power export pays off uner a \textit{ToU} tariff, which results in the least annual electricity cost for consumers, but this might not be most beneficial for DNSPs. Generally, we can conclude that customers are likely to be indifferent between these tariff types, since the annual costs values are quite close.

\subsection{Daily and Monthly Peak Demand} \label{daily_monthly}

\par
\begin{figure}[t]
	\centering
	\includegraphics[scale = 0.65]{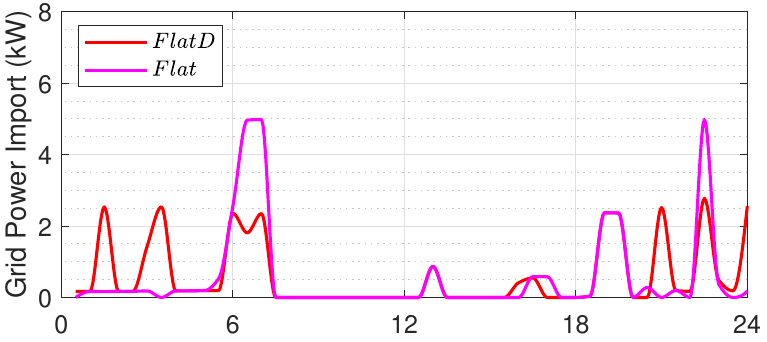}
	\includegraphics[scale = 0.65]{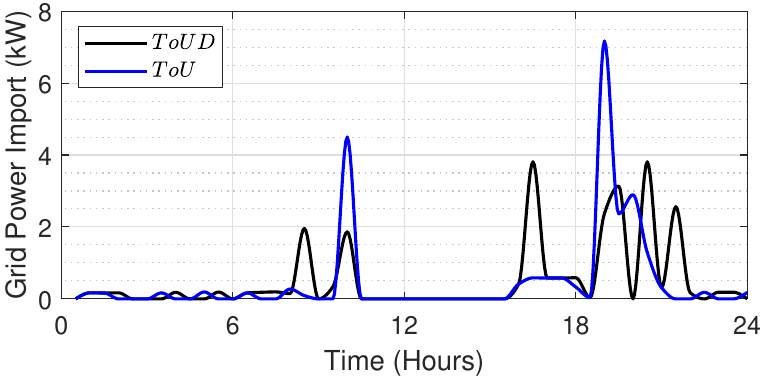}
	\caption{Illustration of peak demand reduction due to the implicit peak demand constraint $\hat{p}$ in the optimisation problem (20).}  
	\label{daily_peak}
\end{figure}

\begin{figure}[t]
	\centering
	\includegraphics[scale = 0.65]{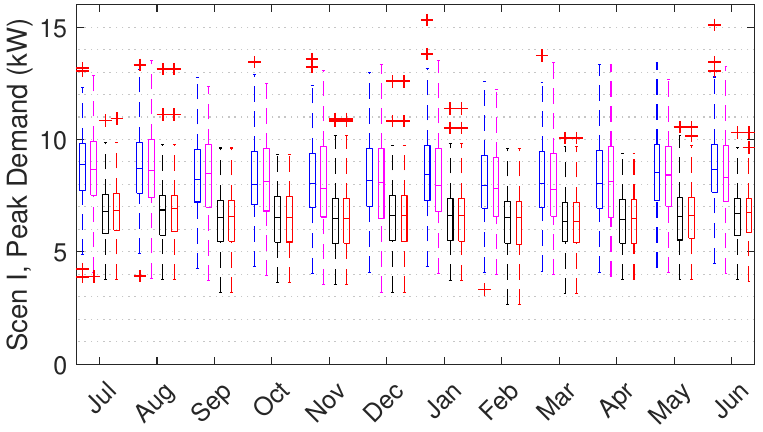}
	\includegraphics[scale = 0.65]{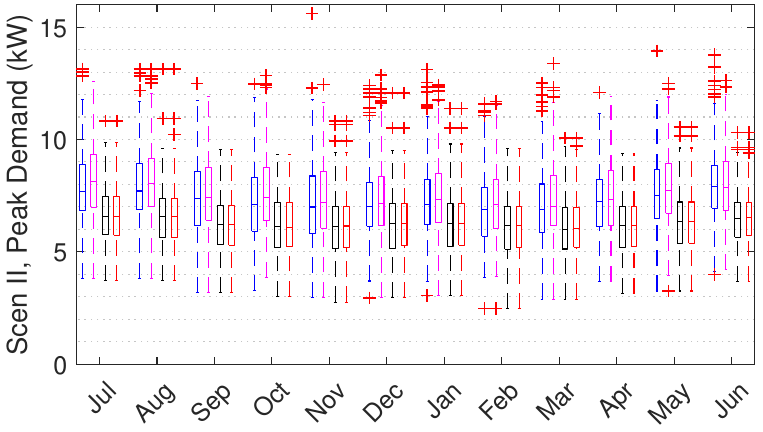}
	\includegraphics[scale = 0.65]{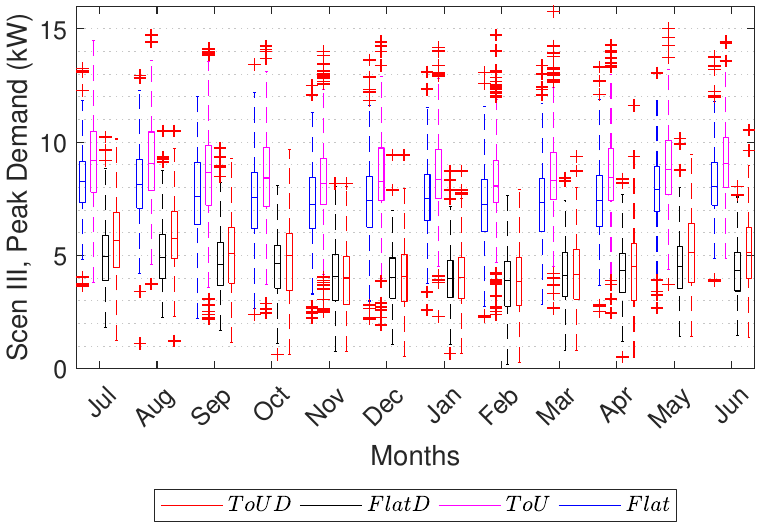}
	\caption{Monthly peak demand of 332 Customers in the three scenarios.} 
	\label{monthly_peak}
\end{figure}

\begin{figure}[t]
	\centering
	\includegraphics[scale = 0.65]{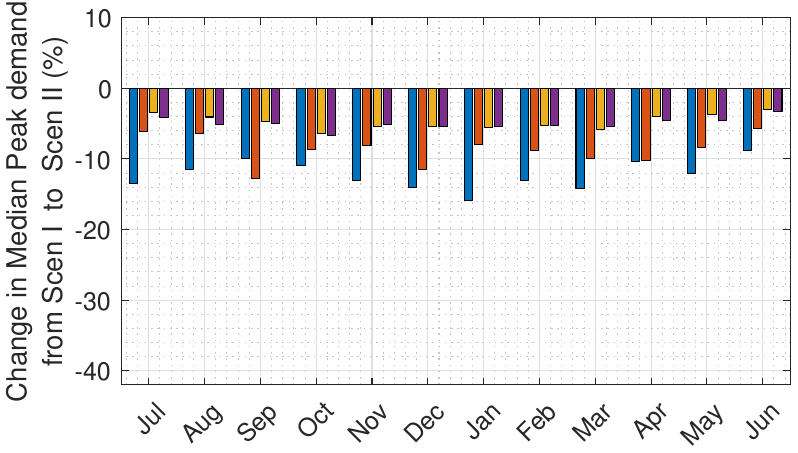}
	\includegraphics[scale = 0.65]{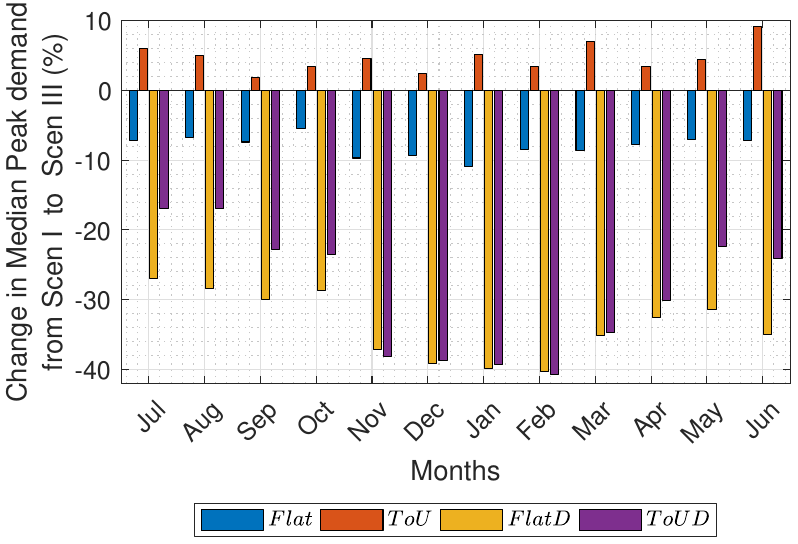}
	\caption{Percentage change in monthly peak demand}  
	\label{peak2}
\end{figure}

 The peak-demand charge has an effect of clipping a customer's daily and monthly grid power import according to~\eqref{eqn22}. Fig.~\ref{daily_peak} illustrates the daily peak demand reduction of Customer 3 (a randomly selected customer) using \textit{demand-based tariffs} (\textit{FlatD} and \textit{ToUD}). We also calculate customers' monthly peak demand under the tariff types by finding the maximum grid import power for each month from the optimisation results. Fig.~\ref{monthly_peak} shows the monthly peak demand for 332 customers in Scenarios I--III while Fig.~\ref{peak2} shows the percentage change in the median peak demand as PV (Scen. II) and PV-batteries (Scen. III) are added.  Generally, using \textit{demand-based tariffs} results in a lower monthly peak demand compared to \textit{energy-based tariffs} due to the additional demand charge to penalize grid power import. 

The results also show that, across all tariff types, solar PV alone (Scen. II) is not sufficient to significantly reduce the peak demand recorded in the base case (Scen. I). Observe in Fig.~\ref{peak2} that solar PV is more effective at reducing the peak demand due to \textit{energy-based tariffs} (up to 16\% with \textit{Flat} tariff in January) than with \textit{demand-based tariffs} (up to 6\% in October). However, with solar PV and batteries (Scen. III), the monthly peak demand even increased (nearly up to 10\% in June) with \textit{ToU} tariff, but was lowered (up to 40\% in February) with \textit{demand-based tariffs} as compared with Scenario I (See Fig.~\ref{peak2}). We can also deduce that ToU-based tariffs perform worst as DER is progressively added compared with flat tariffs (\textit{Flat} and \textit{FlatD}). This is due to the creation of new peaks when all batteries charge at off-peak times to minimise customers' electricity costs.

\begin{figure*}[tbh!] 
\centering
	\hspace{-0.8em}
	\subfloat{%
		\includegraphics[scale = 0.65]{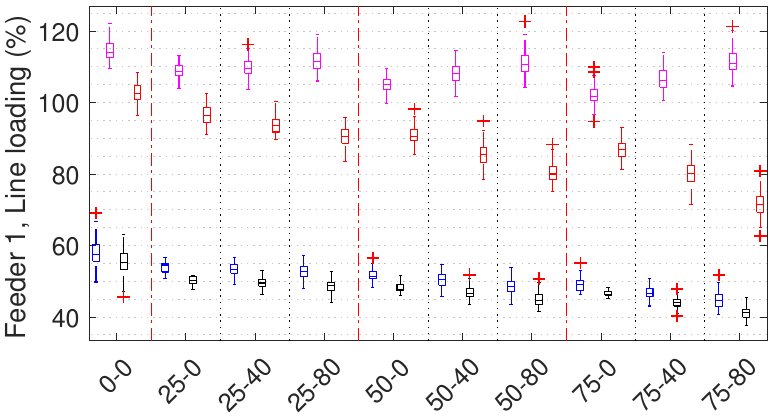}  
	}  \hspace{1.3em}
	\subfloat{%
		\includegraphics[scale = 0.65]{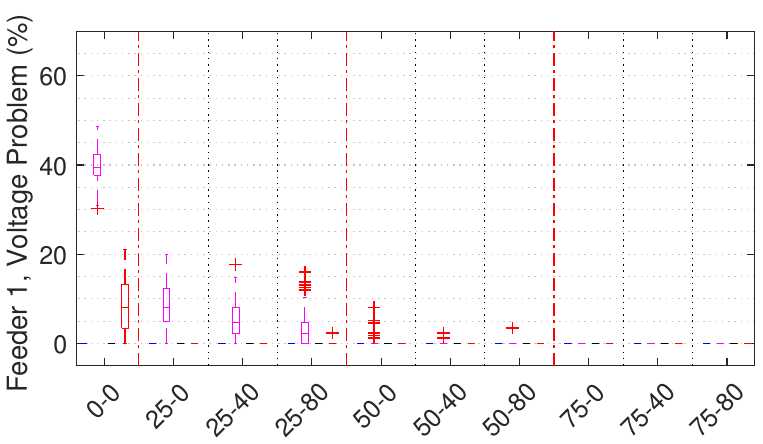} 
	} 
	
	\subfloat{%
		\includegraphics[scale = 0.65]{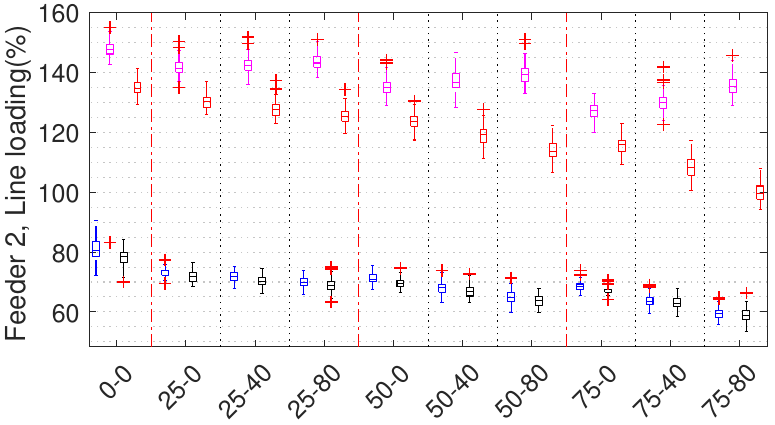} 
	} \hspace{1.2em}
	\subfloat{%
		\includegraphics[scale = 0.65]{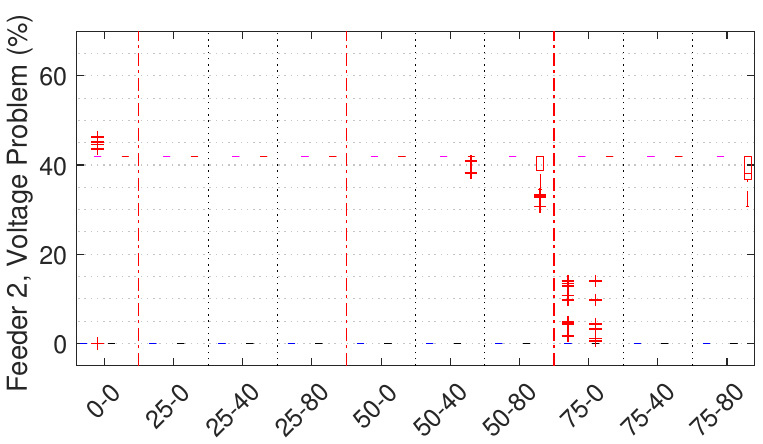} 
	}
	\renewcommand{\thesubfigure}{a}
	\subfloat[]{
	\includegraphics[scale = 0.65]{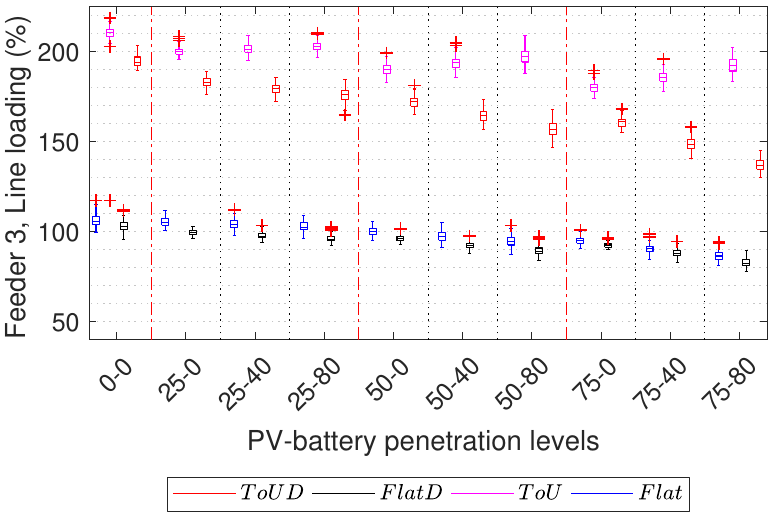} \label{loading_voltage_profiles_a} 
	} \hspace{1em}
	\renewcommand{\thesubfigure}{b}
	\subfloat[]{%
		\includegraphics[scale = 0.65]{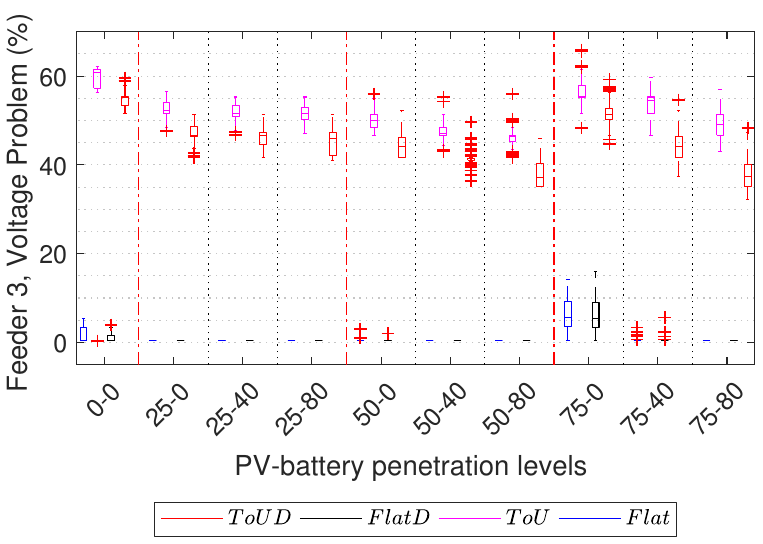} \label{loading_voltage_profiles_b}
	}
	\caption{(a) Feeder head loading level and (b) percentage of customers with voltage problems for Feeder 1 (top), Feeder 2 (middle) and Feeder 3 (bottom). The black dotted lines separate the battery ownership levels (of 0, 40 and 80\%) at each PV penetration level (of 0, 25, 50 and 75\%).}
	\label{loading_voltage_profiles}
\end{figure*}

\subsection{Effects of Network Tariffs on Line Loading} \label{line_loading}
In this section, we analyse the feeder head loading for the different PV-battery penetration levels (Fig.~\ref{loading_voltage_profiles_a}). The loading levels are generally high because we have shown the phases with the highest loading (other phases follow similar pattern) for each feeder and also examined the maximum feeder head loading over the year for each MC simulation. The results show that \textit{ToU} tariff perform worst as the battery penetration level increases, which is in conformity with the results in~\cite{pimm2018time}. This is due to the  batteries' response to ToU pricing by charging at off-peak times, thereby creating new peaks. Furthermore, ToU-based tariffs (\textit{ToU} and \textit{ToUD}), can adversely affect line loading due to large grid imports at off-peak times and reverse power flows resulting from power export. This can be mitigated by adding a demand charge (\textit{ToUD}) to at least clip the grid import levels, with the aid of batteries. As observed, line loading increased with higher battery penetration with \textit{ToU} tariff, while it reduced with \textit{ToUD} tariff. Contrarily, \textit{Flat} tariff results in lower line loading for all feeders. By including a demand charge to the flat tariff (\textit{FlatD}), line loading is reduced even further as seen in all three feeders. This works well with increasing battery penetration in both fairly balanced (Feeders 1 and 2) and unbalanced LV networks (Feeder 3) since there are no incentives for large grid power exports as with ToU tariffs.

\subsection{Effects of Network Tariffs on Customer Voltage Level} \label{voltage_profile}
In terms of customer voltage profiles, Fig.~\ref{loading_voltage_profiles_b} shows that \textit{ToU} tariff results in higher voltage problems in all three feeders compared to the other tariffs. This is particularly obvious in the case of the unbalanced feeder (Feeder 3), but can be mitigated by adding a demand charge to the ToU tariff (\textit{ToUD}). In this case, batteries are useful in reducing voltage problems. \textit{Flat} tariff, on the other hand, performs better than ToU-based tariffs in keeping customer voltage at the right levels. And again, by adding a demand charge to the flat tariff (\textit{FlatD}), there is a slight improvement in the customer voltage profiles.

\section{Conclusions and further work} \label{conclusion}
\par In this research, we have shown that in the presence of DER, adding a peak demand charge to either a Flat or ToU tariff effectively reduces peak demand and subsequently line loading.
\par To reduce a customer's peak demand, we have proposed a computationally efficient optimisation formulation that avoids the computationally expensive min-max formulation used in alternative approaches. We have demonstrated that the novel formulation, which can be seamlessly integrated into a customer’s HEMS, can be used in conjunction with DER-specific tariffs to achieve better network management and cost-reflective network charges.

Generally, flat tariffs perform better than ToU tariffs for mitigating voltage and alleviating line congestion problems. We conclude that, in the context of reducing network peaks, flat tariffs with a peak demand charge will be most beneficial for DNSPs. 
With respect to customer economic benefits, the best tariff depends on the amount of DER a customer possesses. However, the cost savings achieved by switching to another tariff type is marginal. Moreover, with reference to our previous work (all customers without EWH)~\cite{azuatalam2017impacts}, we can also conclude that the EWH has equal impacts across all tariff types in terms of line loading. However, with EWH, the line loading is generally higher.

\par In this study, we have not explicitly tested these tariffs for cost-reflectivity, although this is implicit in the results. In this regard, our next task will focus on the design of these tariffs using established principles in economic theory rather than using already published tariffs from DNSPs.

\bibliographystyle{IEEEtran}
\small{\bibliography{generic-color}}

\end{document}